\documentclass[]{pasj01}
\draft
\begin{document} 
\Received{}
\Accepted{}
\title{X-ray emission from the mixed-morphology supernova remnant HB 9}

\author{
Mariko \textsc{Saito}\altaffilmark{1},
Shigeo \textsc{Yamauchi}\altaffilmark{1$\ast$}, 
Kumiko K. \textsc{Nobukawa}\altaffilmark{1},
Aya \textsc{Bamba}\altaffilmark{2, 3},
and Thomas G. \textsc{Pannuti}\altaffilmark{4}
}
\altaffiltext{1}{Faculty of Science, Nara Women's University, Kitauoyanishimachi, Nara 630-8506, Japan}
\email{yamauchi@cc.nara-wu.ac.jp}
\altaffiltext{2}{Department of Physics, The University of Tokyo, 7-3-1 Hongo, Bunkyo-ku, Tokyo 113-0033, Japan}
\altaffiltext{3}{Research Center for the Early Universe, School of Science, The University of Tokyo, 7-3-1 Hongo, Bunkyo-ku, Tokyo 113-0033, Japan}
\altaffiltext{4}{Department of Physics, Earth Science and Space Systems Engineering, Morehead State University, 235 Martindale Drive, Morehead, KY 40351, USA}
\maketitle

\begin{abstract}
We present the results of a spectral analysis of the central region of the mixed-morphology supernova remnant HB 9. 
A prior Ginga observation of this source detected a hard X-ray component above 4 keV and 
the origin of this particular X-ray component is still unknown. 
Our results demonstrate that the extracted X-ray spectra are best represented by a model 
consisting of a collisional ionization equilibrium plasma 
with a temperature of $\sim$0.1--0.2 keV (interstellar matter component) 
and an ionizing plasma with a temperature of $\sim$0.6--0.7 keV and an ionization timescale 
of $>$1$\times$10$^{11}$ cm$^{-3}$ s (ejecta component). 
No significant X-ray emission was found in the central region above 4 keV. 
The recombining plasma model reported by a previous work does not explain our spectra.
\end{abstract}

\section{Introduction}

HB 9 (G~160.9$+$2.6) is a supernova remnant (SNR) having a shell-like morphology with a  large angular extent of 
$\sim$2$^{\circ}$ in the radio band. 
The measured integrated flux density at 1 GHz and radio spectral index
$\alpha$ (defined as $S$ $\propto$ $\nu$$^{-\alpha}$) are 110 Jy and 0.64, respectively
\citep{Leahy1991,Leahy1998,Leahy2007}.
Based on observations of H\emissiontype{I} cloud structures which are probably associated with HB 9, 
a distance to HB 9 of only 0.8$\pm$0.4 kpc has been proposed \citep{Leahy2007}. 
\citet{Sezer2019} reported a consistent result of 0.6$\pm$0.3 kpc. 
The age was estimated to be 4000--7000 yr \citep{Leahy2007}, and hence HB 9 is a middle-aged SNR. 
\citet{Araya2014} reported the detection of extended gamma-ray emission from the position of HB 9 based on analysis of 
5.5 yrs of data collected by Fermi observations. 
Those authors derived fits to the gamma-ray spectrum using both leptonic and hadronic models. 
More recently, through analysis of 10 yrs of data collected by Fermi observations, 
\citet{Sezer2019} discovered three gamma-ray point sources seen toward the position of HB 9. 
Of these three sources, two are located within the extended gamma-ray emission region and the radio shell. 

X-ray observations of HB 9 were made by Einstein \citep{Leahy1987}, Ginga (\cite{Yamauchi1993}, hereafter YK93), 
and ROSAT \citep{Leahy1995}. 
The X-ray imaging observations showed that the X-ray morphology is 
center-filled: the contrast between the X-ray morphology and its shell-like radio morphology establishes 
HB 9 as a mixed-morphology (MM) SNR \citep{Rho1998}. 
Prior X-ray spectral analyses of this SNR have revealed the existence of a thin thermal plasma associated with HB 9. 
The initial analysis of data from the Einstein observations 
suggested the presence of electron temperature ($kT_{\rm e}$) variations from approximately
$kT_{\rm e}$$\sim$0.4 to $\sim$1.2 keV, with the temperature decreasing from center to edge \citep{Leahy1987}.
However, later analysis of data from the ROSAT observations \citep{Leahy1995} found no evidence for significant temperature 
variations across the X-ray emitting plasma. 
The values for $kT_{\rm e}$ reported by those authors for different portions of the plasma only spanned the range of $kT_{\rm e}$$\sim$0.7--0.8 keV. 

The large area counter (LAC) onboard Ginga was a non imaging detector with a field of view (FOV) of \timeform{1D}$\times$\timeform{2D} (FWHM). 
YK93 reported the presence of a hard X-ray component in the LAC spectrum in addition to the soft thermal emission. 
The temperature of the soft X-ray component was estimated to be $kT_{\rm e}$$\sim$0.4--0.7 keV, 
while the hard X-ray component was fitted with either a thin thermal plasma model with a higher temperature ($kT_{\rm e}$$\sim$6--7 keV) 
or a power law function with a photon index of $\Gamma$$\sim$2.3. 
The origin of this hard X-ray component is not clear. 
YK93 speculated about different mechanisms that might produce it: 
the background active galactic nucleus (AGN) known as 4C 46.09 seen in projection toward HB 9 
($z$=0.195: \cite{Seward1991}), an intracluster gas associated with the cluster of galaxies to which 4C 46.09 belongs, 
or a second X-ray emitting plasma associated with HB 9.
The authors argued that 4C 46.09 and the intracluster gas cannot explain the total flux. 
If the hard X-ray component is associated with HB 9, it is another issue how the hard X-ray component was made. 
Unfortunately, HB 9 has not yet been the target of pointed observations made by the current
generation of X-ray observatories and a clear understanding of the properties of the hard X-ray component and X-ray emitting
plasma (including the true nature of the hard X-ray component) remains elusive.

Recent X-ray observations have revealed that some fraction of MM SNRs have a recombining plasma (RP) which has a higher ionization state than that expected from the electron temperature 
(e.g., IC 443: \cite{Yamaguchi2009}, W49B: \cite{Ozawa2009}, W28: \cite{Sawada2012}, G346.6$-$0.2: \cite{Yamauchi2013}). 
\citet{Sezer2019} analyzed the Suzaku data of HB 9 and reported that a spectrum of a part of the central region of the SNR is represented by the RP model. 

To improve our understanding of HB 9 -- in particular, its X-ray spectral properties and the nature of 
the reported hard X-ray component -- 
we reanalyzed the Suzaku data collected from observations made of the central extended X-ray emission of this SNR.
In this paper, we report on the results of the spectral analysis. 
The quoted errors are the 90\% confidence level unless otherwise mentioned. 

\section{Observations}

\begin{figure}
  \begin{center}
\includegraphics[width=8cm]{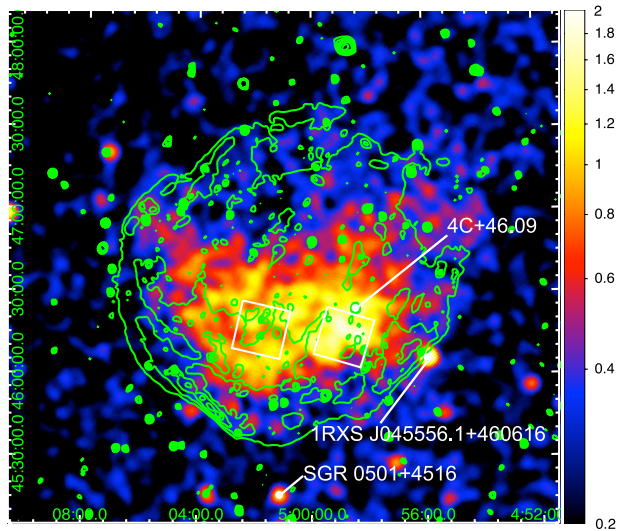} 
  \end{center}
  \caption{
  X-ray (ROSAT, color) and radio band (1420 MHz, processed by the Canadian Galactic Plane Survey Consortium, green contour) images 
  of HB 9, taken from the catalog of High Energy Observations of Galactic Supernova Remnants (\cite{Ferrand2012}; http://www.physics.umanitoba.ca/snr/SNRcat). 
  Squares show the XIS FOV of the present observations. The scale is an arbitrary unit. 
}\label{fig:img}
\end{figure}

Suzaku observations of the SNR HB 9 were carried out with the X-ray Imaging Spectrometor (XIS, \cite{Koyama2007}) placed at the focal planes 
of the thin foil X-ray Telescopes (XRT, \cite{Serlemitsos2007}). 
Two positions were observed: east and west parts of the central bright region of HB 9. 
The XIS FOVs (\timeform{17.'8}$\times$\timeform{17.'8}) are shown on the ROSAT intensity map in figure 1.
The data used in this analysis are listed in table 1.  

The XIS consisted of 4 sensors: XIS sensor-1 (XIS1) is a back-side illuminated CCD (BI), while
the other three XIS sensors (XIS0, 2, and 3) are a front-side illuminated CCD (FI).
Since XIS2 turned dysfunctional in 2006 November, observations were made with the XIS0, 1, and 3. 
A small fraction of the XIS 0 area was not used because of damage, possibly due to the impact of a micrometeorite on 2009 June 23.
The XIS was operated in the normal clocking mode.
The XIS employed the spaced-row charge injection (SCI) technique to rejuvenate 
its spectral resolution by filling the charge traps with artificially injected electrons through CCD readouts.
Details concerning on the SCI technique are given in \citet{Nakajima2008} and \citet{Uchiyama2009}.

\begin{table*}[t] 
 \caption{Observation logs.}
  \centering
   \begin{tabular}{lcccc} \hline
    Part & Obs. ID & Obs. date & (R.A., Dec.)$_{\rm J2000.0}$ & Exposure \\ 
      &              & start time -- end time &                                                 & (ks) \\\hline
     West & 509032010 & 2014-09-30 19:02:01 -- 2014-10-01 22:00:16 & (4:58:49, 46:13:56) & 49.8  \\
     East &  509033010 & 2014-09-29 16:20:36 -- 2014-09-30 19:00:16 & (5:01:45, 46:16:46) & 51.1  \\ 
 \hline
 \end{tabular}
\end{table*}

\section{Analysis and Results}

Data reduction and analysis were made using the HEAsoft version 6.25, XPEC version 12.10.1, and the AtomDB version 3.0.9.
The XIS data in the South Atlantic Anomaly, during the earth occultation, and at the low elevation angle from the earth rim of $<5^{\circ}$ 
(night earth) and $<20^{\circ}$ (day earth) were excluded. 
Removing hot and flickering pixels, the data with grades 0, 2, 3, 4, and 6 were used. 
The XIS pulse-height data were converted to 
Pulse Invariant (PI) channels using the {\tt xispi} software 
and the calibration database version 2018-10-10. 
The resultant exposure times are listed in table 1.

During the observations, count rates of the non-X-ray background (NXB) of XIS 1 were systematically higher than those of the NXB data 
generated by {\tt xisnxbgen} \citep{Tawa2008}. 
Accordingly, we utilized only the FI in the following analysis.  

\subsection{Image}

\begin{figure*}
  \begin{center}
\includegraphics[width=14cm]{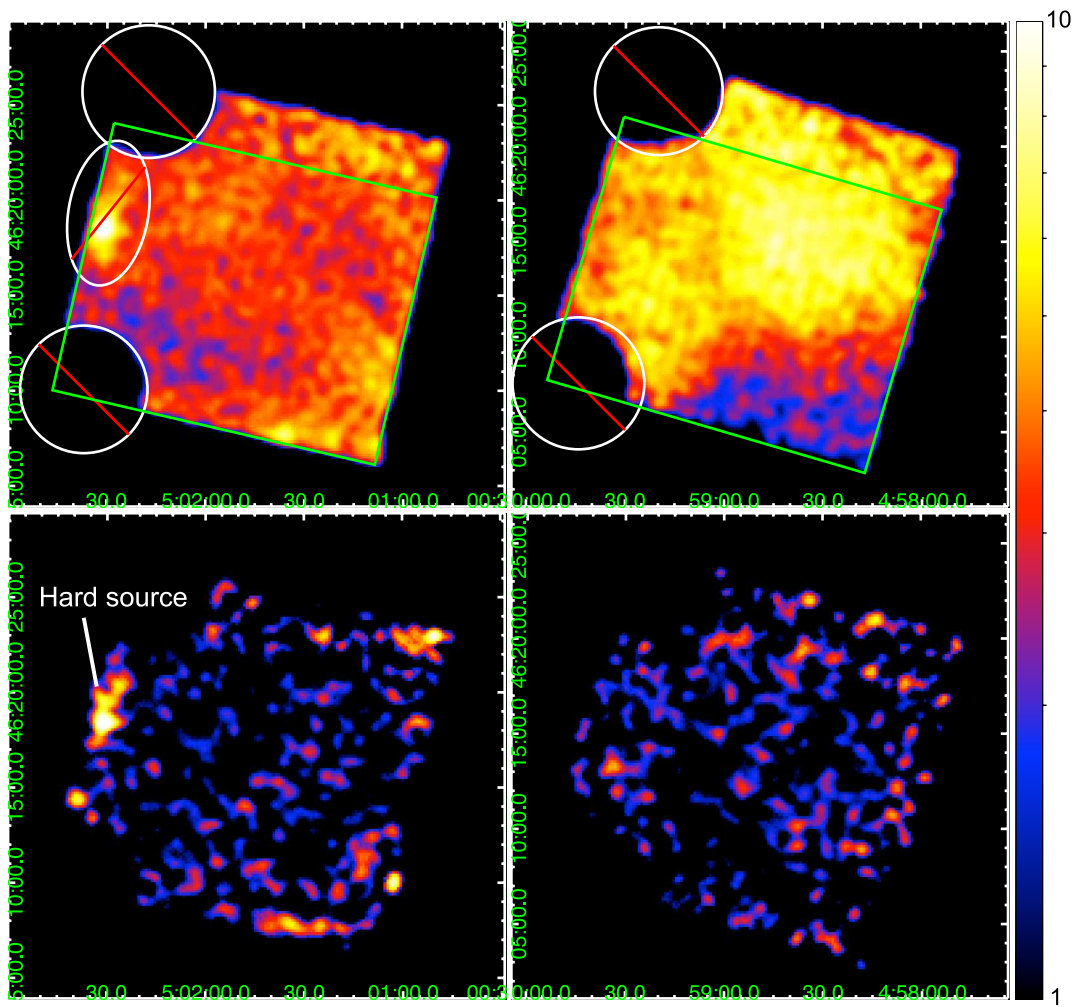} 
  \end{center}
  \caption{XIS images of HB 9: upper panels are soft X-ray band images in the 0.5--2 keV and lower panels are hard X-ray band images in the 4--10 keV.
  Left is the east part and right is the west part. 
  The NXB subtraction and the vignetting correction were made. The coordinates are J2000.0. 
  The color bar shows intensity levels in the arbitrary unit. 
  The green square excluding the areas by the white lines shows the region from which the X-ray spectrum is extracted.
}\label{fig:img}
\end{figure*}

Figure 2 shows X-ray images in the soft X-ray (0.5--2 keV) and hard X-ray (4--10 keV) bands. 
The soft X-ray band images are consistent with those of ROSAT (see figure 1). 
In contrast, the hard X-ray band images show no significant X-ray emission except for a point source at the east edge of the FOV of the east part (hereafter the hard source). 
The position of the hard source was estimated to be (RA, Dec)$_{\rm J2000.0}$=(\timeform{5h02m31.0s}, \timeform{+46D18'29''}) with an uncertainty of $\sim$1$'$.  
Using the SIMBAD database, we identified the source TYC 3344-311-1=2MASS J05023446$+$4618385 
(with an offset of 37$''$) as a possible counterpart to the hard source.
The spectral and temporal properties of the hard source are presented in the Appendix.

\subsection{Spectrum}

The spectrum of the sky background was estimated from the the anti-center sky in the similar Galactic latitude to that of HB 9 
[Obs ID 409019010, ($l$, $b$)=(\timeform{196.D96}, \timeform{+1.D53})].
The spectrum consists of the Milky way halo (MWH), local hot bubble (LHB) and cosmic X-ray background (CXB). 
Since HB 9 is located at the anti-center region with the Galactic latitude of $b$=\timeform{2.D6}, 
the Galactic ridge X-ray emission (GRXE) is ignored \citep{Uchiyama2013,Yamauchi2016}. 
The parameters of the MWH and LHB were assumed to be the same as the values used in \citet{Hirayama2019}, 
while those of the CXB were fixed to the values in \citet{Kushino2002}. 
In addition, after 2011, the Suzaku spectrum exhibited an emission line at 525 eV, 
which is a fluorescent line from atmospheric Oxygen (O\emissiontype{I}), originating from solar X-rays \citep{Sekiya2014}. 
Thus, we also added a Gaussian representing the O\emissiontype{I} line at 525 eV in the spectral fitting. 
Response files, redistribution matrix files (RMFs), and ancillary response files (ARFs) were made using {\tt xisrmfgen} and {\tt xissimarfgen} \citep{Ishisaki2007}, respectively. 
Since the ARFs are normalized to the flux from the assumed emission region, 
the ARFs for the sky background were made assuming the same uniform sky as in the background model fit in \citet{Hirayama2019}, 
while those for the SNR component were made using the sky image obtained with the Suzaku XIS.

\begin{figure*}
  \begin{center}
\includegraphics[width=8cm]{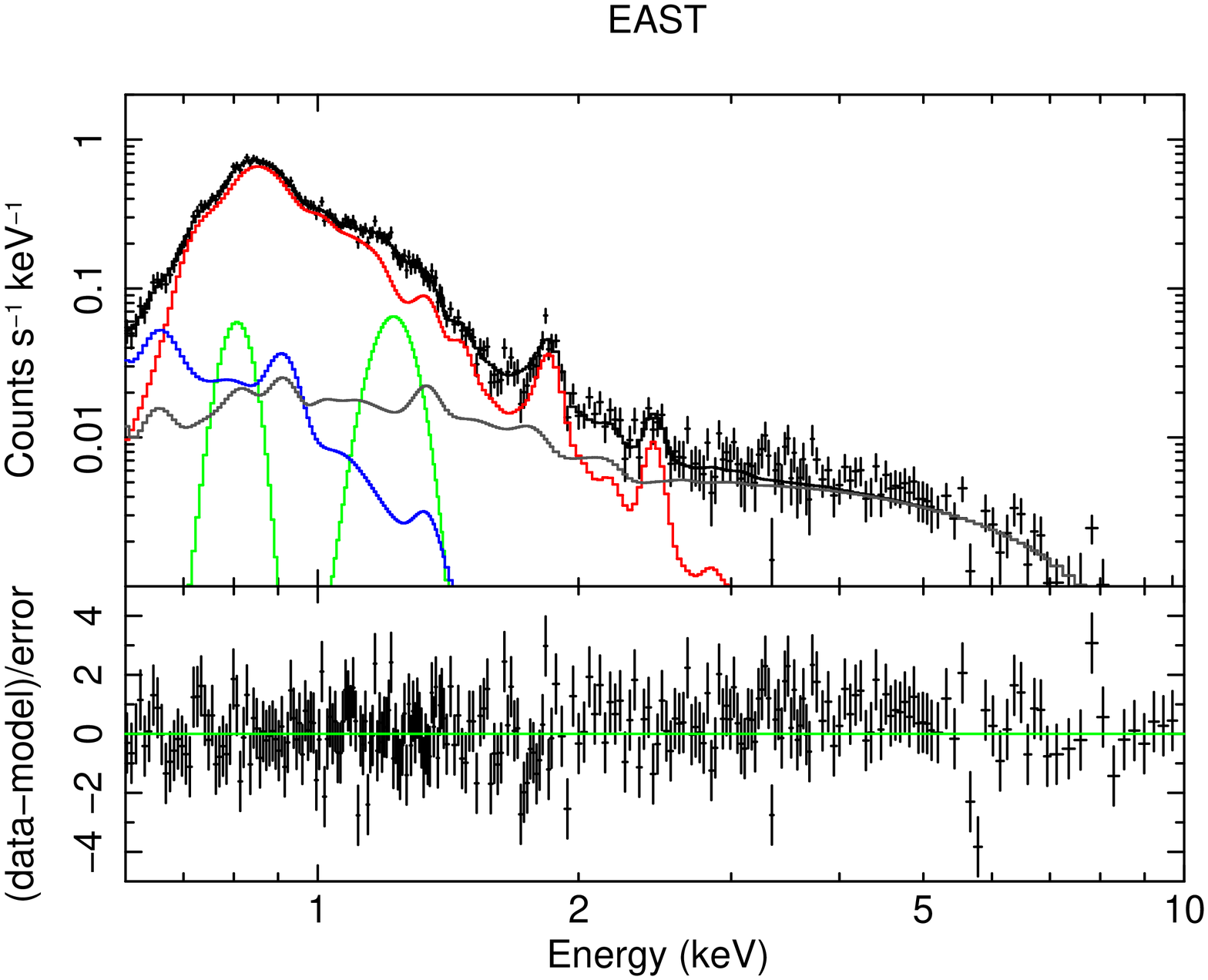} 
\includegraphics[width=8cm]{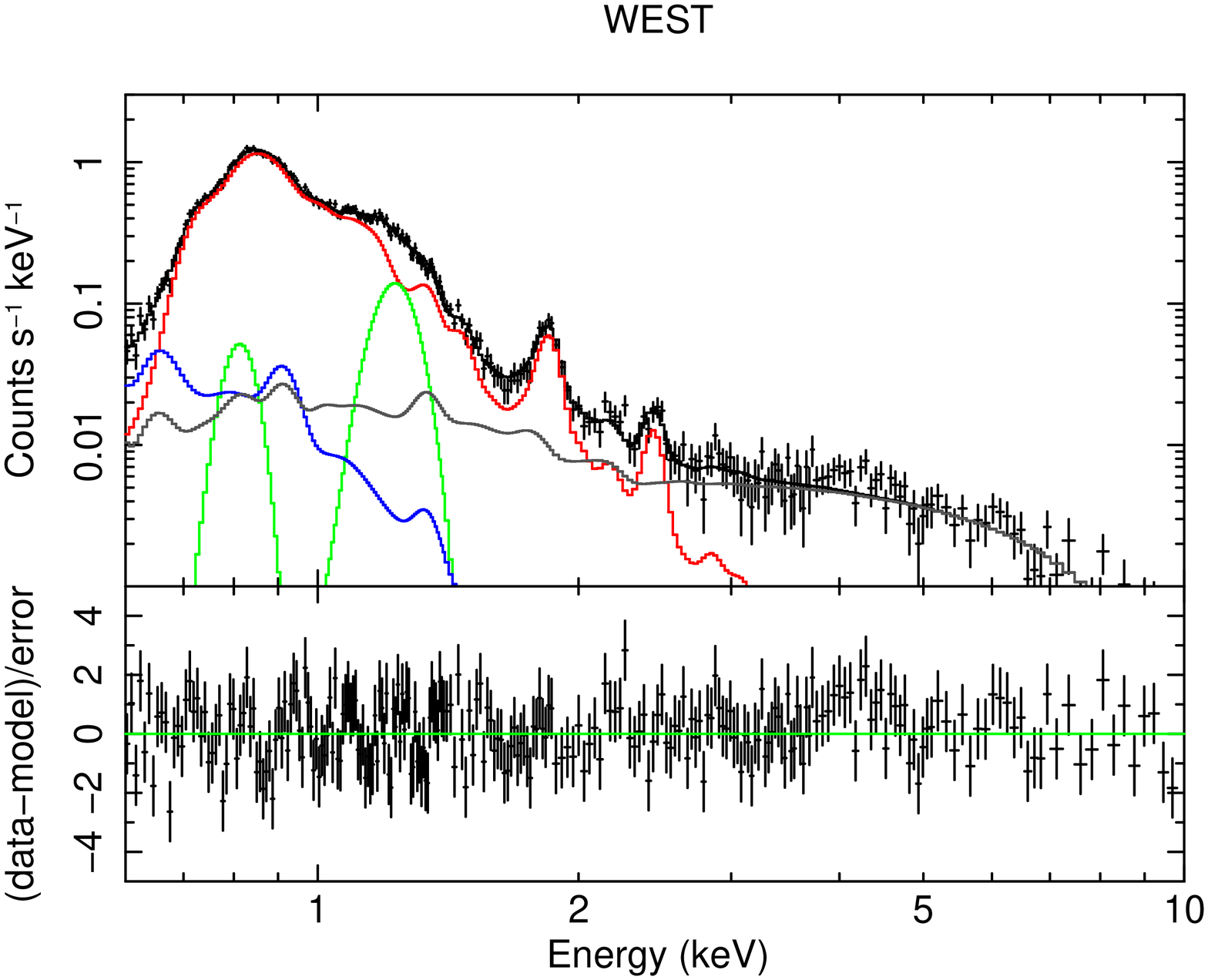} 
  \end{center}
  \caption{XIS spectra (upper panel) and residuals from the best-fit model (lower panel): Left is the east part and right is the west part. 
  The blue, red, green, and gray solid lines show emission from ISM, ejecta, Fe-L lines, and the sky background (MWH$+$LHB$+$CXB), respectively. Errors of data points are at the 1$\sigma$ level. 
}\label{fig:img}
\end{figure*}

Spectra of HB 9 were extracted from the regions shown in figure 2. 
The regions corresponding to the calibration source, the hard source, and damaged pixels were excluded. 
The NXB estimated using {\tt xisnxbgen} \citep{Tawa2008} was subtracted. 
In order to increase photon statistics, we added the spectra of XIS 0 and 3 and  
re-binned the spectra with $\gtsim$25 counts for $<$2 keV and $\gtsim$10 counts for $>$2 keV. 
Figure 3 shows the NXB-subtracted spectrum in the 0.6--10 keV band. 
The gray line shows contribution of the sky background. 
It clearly shows that X-ray emission from the SNR is detected only below 3 keV, which is consistent with the image (section 3.1 and figure 2). 

Taking account of the low electron density of the interstellar medium (for example, $\sim$1 cm$^{-3}$) and the age of SNRs (10$^{3-4}$ yr), most SNRs would be in an ionizing phase. 
Thus, we applied an ionizing plasma (IP) model ({\tt vrnei} model in XSPEC) with low-energy absorption.  
The initial electron temperature, $kT_{\rm init}$, was fixed to 0.0808 keV (the minimum value of the code).
The cross section of the photoelectric absorption and the abundance tables were taken from \citet{bcmc1992} and \citet{Anders1989}, respectively. 
Abundances of Ne, Mg, Si, S, and Fe (=Ni) were free parameters and the others were assumed to be solar. 
We added two Gaussians at $\sim$0.8 keV and $\sim$1.2 keV reproducing features due to  
incomplete atomic data for the Fe-L shell complex in the current plasma model (e.g., \cite{Nakashima2013}). 
In the spectral fitting, we applied the energy scale adjustment using a linear function. 
Although the model represented the spectrum above $\sim$1 keV ($\chi^2$/d.o.f.=367/257 and 347/257 for the east and west parts, respectively), 
we found some residuals less than 0.7 keV. 
Thus, we added another plasma model in the collisional ionization equilibrium (CIE) state ({\tt apec} model in XSPEC). 
Since the abundance of the CIE model was not well constrained, the abundance was fixed to the solar value.  
The model well represented the spectra: the $\chi^2$/d.o.f. value is 315/255 for the east part and 281/255 for the west part.
The electron temperatures of the CIE plasma and IP components are $\sim$0.1--0.2 keV and $\sim$0.6--0.7 keV, respectively. 
The best-fitting model is plotted in figure 3 and the best-fitting parameters are listed in table 2. 

We note that the intensity of the sky background varies by $\pm$10 \% and then the results are the same within the errors.
We also carried out the spectral analysis using another sky background data obtained at the anti-center 
[Obs ID 703019010, ($l$, $b$)=(\timeform{165.D08}, \timeform{+5.D70})], and obtained consistent results. 
Taking account of variations in errors derived from different sky background data, we estimated uncertainties of spectral parameters (table 2). 

\begin{table*}
\caption{The best-fitting spectral parameters for HB9.}\label{tab:first}
\begin{center}   
 \begin{tabular}{lcccc}
      \hline
       Parameter 	& \multicolumn{4}{c}{Value$^{\ast}$}  \\
                         & \multicolumn{2}{c}{East} & \multicolumn{2}{c}{West}\\
     \hline 
     \multicolumn{5}{c}{Absorption} \\
       $N_{\rm H}$ ($\times$10$^{21}$ cm$^{-2}$) 	&  1.6$^{+0.5}_{-0.4}$ & (0.2--2.1) 	& 2.0$^{+0.3}_{-0.4}$ &(1.3--2.3)\\
     \multicolumn{5}{c}{ISM: Collisional ionization equilibrium plasma} \\
         $kT_{\rm e}$ (keV)			& 0.15$^{+0.03}_{-0.02}$ & (0.13--0.19)			&  0.15$^{+0.04}_{-0.02}$ &(0.13--0.20) \\
         Abundance$^{\dag}$			& 1 (fixed) & 					& 1 (fixed) & \\
     \multicolumn{5}{c}{Ejecta: Ionizing plasma} \\
        $kT_{\rm e}$ (keV)			& 0.67$\pm$0.02 & (0.65--0.71)				&  0.66$\pm$0.01 & (0.65--0.68) \\
	$n_{\rm e} t$ ($\times$10$^{11}$ cm$^{-3}$ s)	& $>$4 & ($>$1)				&  $>$5 & ($>$2)	\\
        Ne$^{\dag}$ 				& 1.5$^{+1.6}_{-0.9}$ &(0.5--4.5)				&  1.5$^{+0.8}_{-0.9}$ &(0.4--3.1) \\
        Mg$^{\dag}$ 				& 1.5$^{+1.5}_{-0.7}$ &(0.7--4.4)				&  2.8$^{+2.3}_{-1.2}$ &(1.3--8.2) \\
        Si$^{\dag}$ 				& 1.8$^{+1.0}_{-0.6}$ &(1.0--4.2)				&  3.6$^{+2.4}_{-1.2}$ &(2.0--9.6) \\
        S$^{\dag}$ 				& 3.1$^{+2.3}_{-0.9}$ &(1.6--5.6)				&  4.9$^{+1.1}_{-1.2}$ &(2.4--8.6) \\
        Fe=Ni$^{\dag}$ 			& 2.8$^{+1.7}_{-1.0}$ &(1.6--6.8)				&  6.4$^{+6.0}_{-2.2}$ &(3.5--15.8) \\
        Others$^{\dag}$ 			&1 (fixed) &  					&  1 (fixed) &   \\
	$\chi^2$/d.o.f.  				& 315/255=1.24 &					&  281/255=1.10 &  \\
   \hline
    \end{tabular}
\end{center}
 $^{\ast}$ Statistical errors and error ranges derived from different sky background data (in parentheses) are quoted. \\
 $^{\dag}$ Abundance relative to the solar value \citep{Anders1989}.\\
 \end{table*}

\section{Discussion}

\subsection{Plasma state}

The X-ray spectrum obtained with Suzaku was well represented by a two-component model of the CIE plasma with a temperature of $\sim$0.1--0.2 keV and 
the IP with a temperature of $\sim$0.6--0.7 keV. 
The former and the latter would be the interstellar matter (ISM) and the ejecta components, respectively. 
The spectrum is well consistent with those of middle-aged SNRs.
 
Recently, \citet{Sezer2019} reported results of the spectral analysis of the Suzaku data.
The spectrum was fitted with a two component model: 
the west part spectrum is represented by the CIE ($kT_{\rm e}$=0.42 keV)$+$RP ($kT_{\rm e}$=1.13 keV and $kT_{\rm init}$=3.14 keV) model, 
while the east part spectrum is explained by the CIE ($kT_{\rm e}$=0.51 keV)$+$IP ($kT_{\rm e}$=0.97 keV) model. 
However, no significant difference in the X-ray spectra between the east and west parts is seen (figure 2 of \cite{Sezer2019}). 

The spectrum in \citet{Sezer2019} is almost the same as ours. 
Using the abundance tables in \citet{Wilms2000} that \citet{Sezer2019} used, 
we applied the CIE$+$RP model with fixing spectral parameters, except for the normalization, to those of \citet{Sezer2019}
to the west part spectrum, and found that the model is completely rejected (a reduced $\chi^2$ value of $>$5).  
In the case of the east part analysis (the CIE$+$IP model fit), 
the electron temperatures in \citet{Sezer2019} are obviously higher than ours. 
The difference between the results in this paper and those given in \citet{Sezer2019} could be due to the differences 
in sky background estimation.

Here, using our spectral data, we examined whether HB 9 possesses an RP or not. 
The CIE$+$RP model of $kT_{\rm init}$=3 keV with freeing electron temperature, recombining timescale, abundances, and $N_{\rm H}$ value was also applied to 
both the east and west spectra,  
but no improvement from the CIE$+$IP model (see table 2) was found. 
The electron temperature and recombining time scale of the ejecta component were 0.6--0.7 keV and $\gtrsim$10$^{12}$ cm$^{-3}$ s, respectively, 
which shows that the plasma is not the apparent RP. 
In fact, the spectra in both the east and west parts show no obvious RP features, 
such as radiative recombination continuum and intense lines from hydrogen-like ions (see residuals in figure 3). 
Thus, the claim that HB 9 possesses the RP has not been established yet.

\subsection{Origin of the hard X-ray component found with Ginga}

YK93 reported the presence of a hard X-ray component in the spectrum observed with Ginga. 
The $N_{\rm H}$-corrected flux of the hard X-ray component in the 2--10 keV band is $\sim$6$\times$10$^{-12}$ erg s$^{-1}$ cm$^{-2}$. 
The authors argued that the hard X-ray component may not be fully attributable to 4C 46.09 and surrounding intracluster gas, and 
at least 30\% of the observed flux ($\gtrsim$2$\times$10$^{-12}$ erg s$^{-1}$ cm$^{-2}$) is likely attributable to HB 9. 
Since the aspect correction was not made, the Ginga flux is a lower limit (YK93). 
We observed the central parts of the HB 9 with Suzaku and could not detect the hard X-ray component 
(the energy flux of $\lesssim$ 3$\times$10$^{-13}$ erg s$^{-1}$ cm$^{-2}$ FOV$^{-1}$ in the 2--10 keV band, if any). 
This indicates that the hard X-ray component would not be a hot interior of HB 9 as proposed in YK93. 

The FOV of the Ginga LAC includes the overall structure shown in figure 1. 
Since several discrete sources have been discovered around HB 9 so far, a substantial flux may be attributable to the sources. 
Here, we calculate the energy fluxes of 3 sources, the hard source found in this observation, a soft gamma-ray repeater SGR 0501$+$4516  \citep{Barthelmy2008}, 
and 1RXS J045556.1$+$460616 \citep{Voges1999}.
The hard source exhibits a hard spectrum, and the observed flux is $\sim$3$\times$10$^{-13}$ erg s$^{-1}$ cm$^{-2}$ in the 2--10 keV band (see Appendix). 
SGR 0501$+$4516 in the quiescent phase was observed with XMM-Newton and 
the spectrum was well fitted with an absorbed blackbody$+$power law model \citep{Camero2014}. 
Using the best-fit parameters, we calculated the flux to be $\sim$2$\times$10$^{-12}$ erg s$^{-1}$ cm$^{-2}$ in the 2--10 keV band.  
1RXS J045556.1$+$460616 is identified with a rotationally variable star TYC 3344-1956-1=2MASS J0455617$+$4606160 (SIMBAD database).
The ROSAT count rate is 0.18 count s$^{-1}$, which corresponds to $\sim$(0.1--1)$\times$10$^{-12}$ erg s$^{-1}$ 
in the 2--10 keV band assuming thin thermal emission with a temperature of 1--3 keV.
Based on the above estimation, most of the hard X-ray component would originate from discrete sources in the Ginga LAC FOV.

\section{Conclusion}

Using data obtained with Suzaku, we conducted a detailed spectral analysis of the central region of the SNR HB 9. The results are summarized as follows.

\begin{itemize}

\item A previous study of HB 9 with the Ginga LAC suggested the presence of a hard X-ray component (YK93). 
We found no significant hard component to the X-ray emission from the central region of HB 9. 
Most of the hard X-ray component would originate from discrete sources in the Ginga LAC FOV.

\item In order to investigate the physical state of the plasma associated with HB 9, 
we fitted the spectrum with a plasma model and found that 
X-ray spectrum was represented by a model consisting of a CIE plasma 
with a temperature of $\sim$0.1--0.2 keV (ISM component) 
and an IP with a temperature of $\sim$0.6--0.7 keV and an ionization timescale of $>$2$\times$10$^{11}$ cm$^{-3}$ s (ejecta component). 
We found no clear evidence of the RP. 

\end{itemize}

\medskip

The present observations only targeted two small parts of the central region of HB 9. 
In order to investigate the overall properties of the SNR, 
observations of the entire region in the wide energy band are required.

\begin{ack}
We would like to express our thanks to all of the Suzaku team. 
The authors wish to thank the referee for constructive comments that improved the manuscript. 
This work is supported in part by
Shiseido Female Researcher Science Grant (AB) and JSPS/MEXT KAKENHI grant numbers 19K03908 (AB) and JP16J00548 (KKN). 
This research has made use of the SIMBAD database, operated at CDS, Strasbourg, France.
\end{ack}

\section*{Appendix: property of the hard source}

A source spectrum was extracted from an elliptical region centered on the source position, while
the background spectrum was extracted from a nearby region in the same FOV. 
We fitted the spectrum after background subtraction with either a power law model or a thin thermal plasma model ({\tt apec} model in XSPEC), modified with low energy absorption. 
The parallax of the optical/infrared counterpart of the source is 0.6438 mas (SIMBAD database), which corresponds to a distance of 1.55 kpc.
Thus, it would be a stellar source. 
Abundances of stellar sources are typically subsolar (0.3--0.5 solar, e.g., \cite{Gudel1999,Baskill2005}). 
Thus, we assumed the abundances to be 0.3 solar. 
The best-fit parameters are listed in table 3. 

The observed energy flux is converted into a luminosity of (0.9--1.0)$\times$10$^{32}$ erg s$^{-1}$. 
The hard spectrum with $\Gamma$ of 1.7 (power law) or $kT_{\rm e}$ of 6.3 keV (apec) and the derived luminosity suggest 
that the hard source is likely to be a cataclysmic variable (e.g., \cite{Ezuka1999,Nobukawa2016}). 

In order to check a time variation, we made a light curve. No significant variation was found during the observation.

\begin{table}[htb]
\caption{The best-fit parameters for the hard source.}\label{tab:first}
\begin{center}   
 \begin{tabular}{lcc}
      \hline
      Parameter 					& \multicolumn{2}{c}{Value} \\
                          				& power law 				& apec \\
      \hline
        $N_{\rm H}$ (cm$^{-2}$) 		&  $<$3$\times$10$^{21}$  	&   $<$2$\times$10$^{21}$ 	\\
        $\Gamma$/$kT_{\rm e}$ (keV) 	& 1.7$^{+0.4}_{-0.2}$ 		& 6.3$^{+3.8}_{-2.1}$ 	\\
        Abundance 				& --- 						& 0.3 (fixed)\\
        Flux$^{\ast}$ 				&  3.4$\times$10$^{-13}$		&  3.2$\times$10$^{-13}$	 \\ \hline
	$\chi^2$/d.o.f.  				& 31.7/28  				& 32.2/28			\\
   \hline
    \end{tabular}
\end{center}
$^{\ast}$ $N_{\rm H}$-corrected energy flux in the 2--10 keV band. The unit is erg s$^{-1}$ cm$^{-2}$. 

 \end{table}


\end{document}